\documentclass[10pt,twocolumn,letterpaper]{article}

\usepackage{iccv}              

\usepackage{multirow}
\usepackage{colortbl}

\definecolor{iccvblue}{rgb}{0.21,0.49,0.74}
\usepackage[pagebackref,breaklinks,colorlinks,allcolors=iccvblue]{hyperref}

\def\paperID{} 
\def\confName{ICCV}
\def\confYear{2025}

\title{Controlling Avatar Diffusion with Learnable Gaussian Embedding}

\author{Xuan Gao \qquad Jingtao Zhou \qquad Dongyu Liu \qquad Yuqi Zhou \qquad Juyong Zhang\thanks{Corresponding author}\\
University of Science and Technology of China\\
}

\begin{document}

\twocolumn[{%
\maketitle
\begin{center}
    \includegraphics[width=\textwidth]{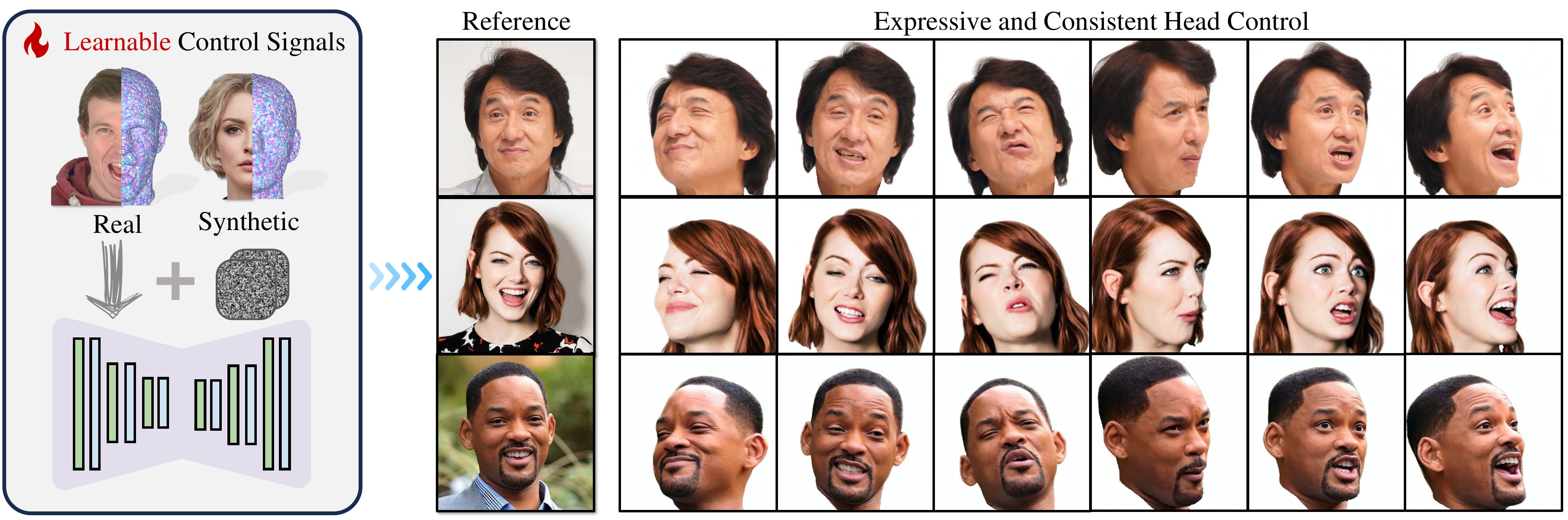}
    \captionof{figure}{We introduce a novel diffusion control signal representation splatted from learnable Gaussians, which is dense, adaptive, expressive, and 3D-consistent. Additionally, we incorporate a real/synthetic token to minimize artifact contamination of the synthetic dataset. Given a single reference image, our model can achieve high-quality, expressive, and consistent head generation.}
    \label{fig:teaser}
\end{center}
}]

{
  \renewcommand{\thefootnote}%
    {\fnsymbol{footnote}}
  \footnotetext[1]{Corresponding Author}
}

\begin{abstract}
Recent advances in diffusion models have made significant progress in digital human generation. However, most existing models still struggle to maintain 3D consistency, temporal coherence, and motion accuracy. A key reason for these shortcomings is the limited representation ability of commonly used control signals(e.g., landmarks, depth maps, etc.). In addition, the lack of diversity in identity and pose variations in public datasets further hinders progress in this area. In this paper, we analyze the shortcomings of current control signals and introduce a novel control signal representation that is optimizable, dense, expressive, and 3D consistent. Our method embeds a learnable neural Gaussian onto a parametric head surface, which greatly enhances the consistency and expressiveness of diffusion-based head models. Regarding the dataset, we synthesize a large-scale dataset with multiple poses and identities. In addition, we use real/synthetic labels to effectively distinguish real and synthetic data, minimizing the impact of imperfections in synthetic data on the generated head images. Extensive experiments show that our model outperforms existing methods in terms of realism, expressiveness, and 3D consistency. Our code, synthetic datasets, and pre-trained models will be released in our project page: \href{https://ustc3dv.github.io/Learn2Control/}{https://ustc3dv.github.io/Learn2Control/}

\end{abstract}    
\section{Introduction}
\label{sec:intro}

Generalizable head modeling has wide applications in AR/VR, film, and entertainment industries. Ensuring high-fidelity rendering while maintaining 3D consistency, temporal consistency, and identity generalization still remains a major challenge. Constructing a powerful head model depends heavily on how the head is represented, how the generative model is trained, and high-quality training datasets.

Early generalizable head models~\cite{blanz1999morphable,cao2013facewarehouse,tran2018nonlinear,li2017learning} generally use meshes to represent the head, but this representation method faces many challenges, such as difficulty in modeling non-facial areas, and unrealistic rendering effects. Some methods use implicit functions ~\cite{hong2021headnerf,sun2023next3d,zhuang2022mofanerf,deng2024portrait4d,deng2024portrait4dv2,tran2024voodooxp,tran2024voodoo,li2023goha} or 3D Gaussians ~\cite{xu2024gphm,chu2024gagavatar} to model the head, further improving the rendering effect of the model. Although these methods have achieved some success, they mainly adopt a one-to-one mapping paradigm, while the relationship between the control signal and the rendered head image is essentially one-to-many. The one-to-one design seriously hinders the expressiveness and generalization ability of the model.

Diffusion models have demonstrated remarkable capabilities in addressing one-to-many problems. Most recent works employ landmarks, normal maps, or depth maps as signals to control the generation of digital humans. Despite some successes achieved by these approaches, we observe that their results still suffer from deficiencies in 3D consistency and expression accuracy. These deficiencies in the generation results are closely related to the control signals used by these diffusion models. For example, the information of landmarks is too sparse, making it hard to capture detailed facial muscle movements or accurately describe head poses. Depth maps and normal maps only represent specific geometric attributes and are inherently camera-dependent, lacking 3D consistency. When such 3D inconsistent control signals are introduced into diffusion, the inconsistency in the generation results is further amplified.

In addition to the representation of control signals, improving data quality and diversity is also crucial for the ability of the trained model. Some studies have utilized game engines to generate synthetic data ~\cite{buehler2024cafca,wang2023rodin,zhang2024rodinhd} to meet the requirements of extensive training data. However, due to the inherent authenticity gap between synthetic data and real data, models trained on these data usually often struggle to produce highly realistic results, or need to be fine-tuned for specific subject to improve the results.

As shown in \Cref{fig:teaser}, in this paper, we introduce a powerful and generalizable diffusion avatar, which is empowered by a learnable Gaussian embedding control signal and a diverse synthetic human head dataset. Different from all existing control signals for digital head generation, we no longer rely on visual cues like landmarks, normal maps, or depth maps, but instead embed a learnable Gaussian field onto the parameterized head surface~\cite{FLAME:SiggraphAsia2017}. This allows the model to adaptively learn and interpret local geometric features. Compared to landmarks, this representation provides a denser spatial representation, and compared to normal or depth maps, it is 3D consistent and conveys richer semantic information. The splatted Gaussian feature map is then injected into a reference-based denoising framework to generate high-quality avatars.

To enhance the model's robustness under different viewpoints, we further introduce a synthetic dataset. Unlike previous methods that rely on game engines for data synthesis, we leverage a state-of-the-art 3D head generative model ~\cite{li2024spherehead} to generate data that closely aligns with the distribution of real human heads. This results in a high-quality dataset with rich identity diversity and a large number of pose variations. In addition, we incorporate a learnable token into the cross-attention module of U-Net, enabling the model to distinguish between generated and synthetic data. This helps prevent artifacts of synthetic data from affecting the realism of the generated results.

In summary, our contributions include the following aspects:

\begin{itemize}
\item We propose a new digital head control signal. Instead of relying on existing visual cues, we leverage feature maps with learnable Gaussian embeddings, significantly enhancing the consistency and authenticity of the generated results.
\item 
To address the limitations of existing public datasets in terms of identity diversity and pose richness, we propose to use synthetic data to improve the ability of the trained model, and use real/synthetic labels to eliminate the impact of artifacts in synthetic data on generated results.

\item Experiments show that our model is significantly better than previous works in terms of expressiveness and consistency. 
It should be noted that the proposed control signal representation and training data synthesis and utilization mechanism is also expected to be extended to more general 3D digital content generation.
\end{itemize}
\section{Related Work}

\label{sec:related}
\subsection{Generalizable Head Avatars}

Mesh-based head models have been extensively studied over the years. Some approaches ~\cite{Khakhulin2022ROME,xu2020deep,retsinas20243d} incorporate neural rendering techniques to enhance face rendering quality. However, due to the inherent limitations of mesh representations, it is still hard to capture detailed facial motions and model non-facial regions
Recent advancements have leveraged NeRF~\cite{mildenhall2020nerf} to achieve photo-realistic head rendering. Some methods integrate parametric facial priors with MLPs~\cite{hong2021headnerf,zhuang2022mofanerf,buehler2024cafca}, tri-planes~\cite{sun2023next3d,chu2024gpavatar,li2023one,li2024generalizable,ma2023otavatar,yu2023nofa,yue2022anifacegan,YeReal3DPortrait}, manifolds~\cite{yue2023aniportraitgan} or volumetric primitives~\cite{cao2022authentic} to enhance controllability and disentanglement within radiance fields. Others learn the disentanglement under the help of some pretrained motion model~\cite{deng2024portrait4d} or directly from  data~\cite{tran2024voodoo,tran2024voodooxp,deng2024portrait4dv2}. Some recent works utilize 3D Gaussians~\cite{kerbl3Dgaussians} as primitives to model generalizable head avatars~\cite{chu2024gagavatar,xu2024gphm}. Another line of works~\cite{yenamandra2021i3dmm,zheng2022imface,giebenhain2023nphm} mainly focus on detailed head geometry. Instead of using radiance field, they use SDF value to accurately represent the surface. 


\begin{figure*}[t]
  \centering
  \includegraphics[width=\linewidth]{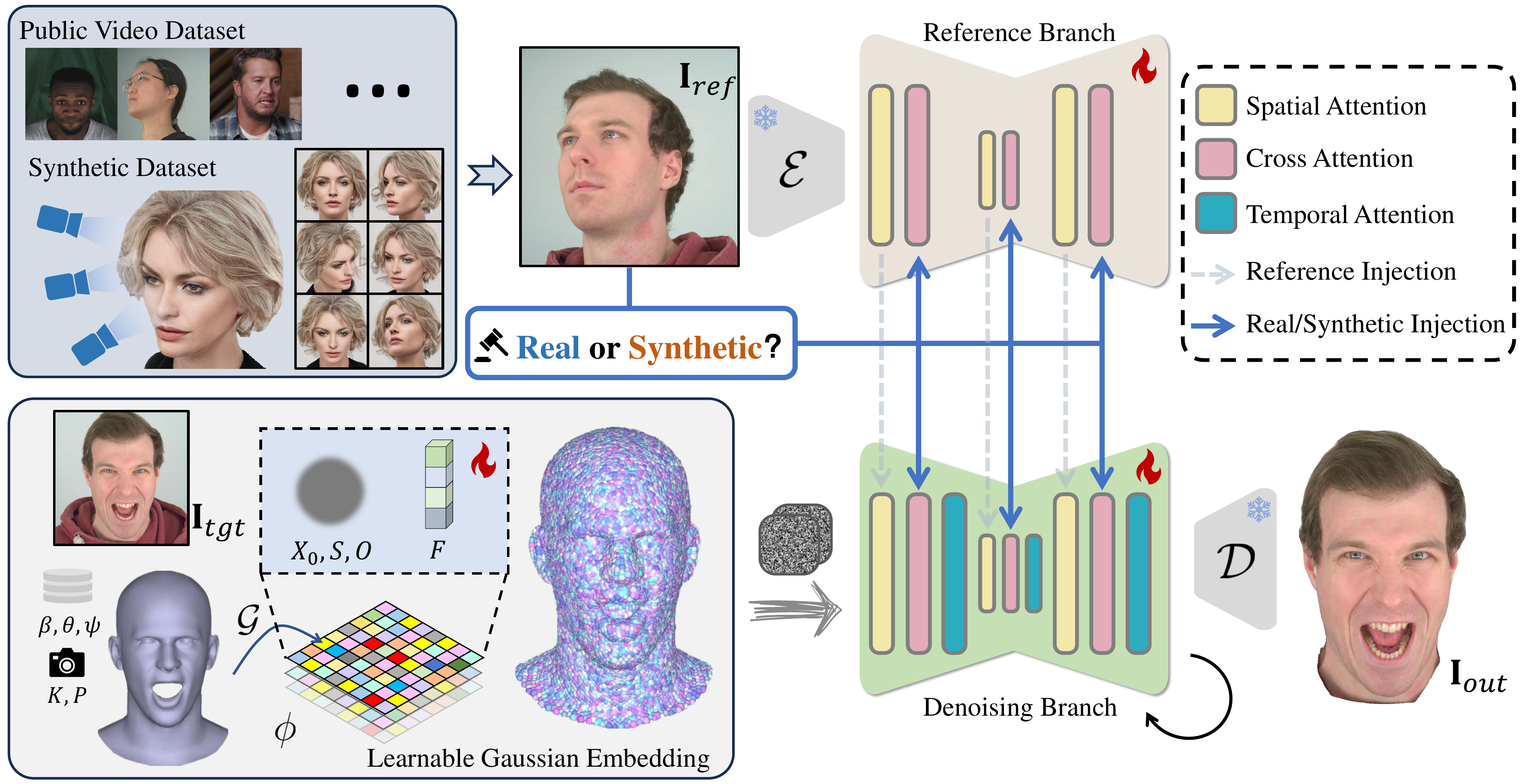}
  \caption{Our pipeline. To address the limitations of existing public datasets in terms of identity diversity and pose richness, we propose to use synthetic data to improve the generalization ability and view consistency of the trained model. We first track the FLAME coefficients of the driving frames $\mathcal{I}_{tgt}$. Then the learnable Gaussians in UV space are transformed to 3D space according to FLAME UV mapping. Subsequently, the transformed Gaussians are projected and splatted to serve as control signals for a reference-guided diffusion model. }
  \label{fig:pipeline}
  \vspace{-2mm}
\end{figure*}
\subsection{Non-diffusion-based Head Animation}
Face animation has been studied for a long time.  Some methods~\cite{Siarohin_2019_NeurIPS, siarohin2021motion, wang2021facevid2vid, mallya2022implicitwarpinganimationimage, zhao2022thinplatesplinemotionmodel, hong2022depth, hong23implicit, guo2024liveportrait} have utilized implicit keypoints as an intermediate motion representation to warp the source portrait. By encoding expressions as latent vectors and injecting them into generator networks, some methods~\cite{zakharov2019fewshotadversariallearningrealistic, Burkov_2020_CVPR, zhou2021pose, wang2022progressivedisentangledrepresentationlearning, drobyshev2022megaportraits, drobyshev2024emoportraits} seek to map driving facial movements onto synthesized portraits via feature-space manipulation. Although these methods have achieved some success, due to the lack of human head priors, they may exhibit some artifacts , particularly under extreme pose changes or large expression variations. Some animation models~\cite{tewari2020stylerigriggingstylegan3d, ghosh2020gifgenerativeinterpretablefaces, ren2021pirenderer, Yin22styleheat, zhang2022sadtalker} integrate 3DMM priors~\cite{blanz1999morphable} into GANs, but their 3D consistency and rendering quality are not satisfying enough.

\subsection{Diffusion-based Head Generation}
Denoising Diffusion Models~\cite{ho2020denoising} have demonstrated remarkable generative capabilities in computer vision. Early works primarily leveraged diffusion models for single-image generation~\cite{paraperas2024arc2face,han2024faceadapter,wang2024instantid} or editing~\cite{brooks2022instructpix2pix,zhao2023diffswap}. FADM~\cite{FADM} was the first diffusion based portrait animation model. More recent works have integrated shared attention mechanisms to inject reference information into the denoising network. Some approaches employ synthetic cross-identity data pairs to enable direct animation driven by real images~\cite{Xportrait,yang2024megactor}, while others utilize landmarks, normal maps, or depth maps as control signals~\cite{prinzler2024jokerconditional3dhead,wei2024aniportrait,ma2024followyouremoji,yang2025showmaker}. Additionally, multi-modal control signal integration~\cite{zhu2024champ,xu2024anchorcrafter,guan2024talkact,zhou2024realisdance} and attention-based region enhancement~\cite{yang2025showmaker} have been explored to improve animation realism and flexibility. There also exist some works that construct the diffusion process in UV space~\cite{lan2023gaussian3diff} or tri-plane space~\cite{wang2023rodin,zhang2024rodinhd}. 
The most closely related works to this paper are DiffusionAvatars~\cite{kirschstein2024diffusionavatars}, ConsistentAvatar~\cite{YangConsistentAvatar} and Stable Video Portraits~\cite{svp}. While all have demonstrated some success in diffusion-based avatar rendering, they require subject-specific training.


Another research direction focuses on leveraging 2D diffusion priors for 3D portrait construction. 
Some methods utilize multi-view diffusion models to ensure view consistency in generated portraits~\cite{kant2025pippo,he2024magicman,tang2024gaf,mendiratta2023avatarstudio,chen2024morphable,Gu_2024_diffportrait3d,lu2025gas}. Others distill image diffusion priors into the construction process~\cite{zhang2023dreamface,gerogiannis2025arc2avatar} for high quality portrait generation.


\section{Method}
\label{sec:method}

Given a reference image $\mathbf{I}_{ref} \in \mathbb{R}^{H\times W\times 3}
$ and a target sequence with length $N$, denoted as 
$\mathcal{I}_{tgt}= \{\mathbf{I}^{n}_{tgt}\in \mathbb{R}^{H\times W\times 3}|n \in [1, N]\}$. 
Our goal is to generate a corresponding output sequence  $\mathcal{I}_{out}=\{\mathbf{I}^{n}_{out}\in \mathbb{R}^{H\times W\times 3}|n \in [1, N]\}$. The generated sequence should faithfully follow the motion in $\mathcal{I}_{tgt}$ while preserving the identity in $\mathbf{I}_{ref}$. 
The overall pipeline of our approach is illustrated in \Cref{fig:pipeline}. We begin with an overview of Stable Diffusion~\cite{rombach2022highresolution} in \Cref{subsec:preliminary}, which serves as the backbone of our approach. In \Cref{subsec:NGHP}, we introduce a novel learnable control signal empowered by learnable Gaussian embedding. Next, in \Cref{subsec:diffusion}, we inject the splatted head features into a reference-guided diffusion model, ensuring that the generated motion aligns with $\mathcal{I}_{tgt}$ while maintaining identity consistency with $\mathbf{I}_{ref}$. To further unleash the potential of our model, in \Cref{subsec:synthetic}, we generate a synthetic multi-view dataset to enhance the identity diversity and view consistency of our model.  

\subsection{Preliminary: Stable Diffusion}
\label{subsec:preliminary}
Stable Diffusion (SD)~\cite{rombach2022highresolution} performs the diffusion process in the latent space rather than directly on pixel space. Given an input image $\mathbf{I}\in \mathbb{R}^{H\times W\times 3}$, the encoder $\mathcal{E}$  maps it into a lower-dimensional latent representation: $\mathbf{z} = \mathcal{E}(\mathbf{I}) \in \mathbb{R}^{h\times w\times channel}$. The decoder $\mathcal{D}$ then reconstructs the image from this latent space:  $\mathbf{I}_{recon} = \mathcal{D}(\mathbf{z})$. 

SD learns to denoise an unstructured Gaussian noise $\epsilon$ to $\mathbf z$ in real image latent domain according to some given condition $c$. During training, the image latent $\mathbf z$ is diffused in $\mathnormal t$ timesteps to produce noise latent ${\mathbf z}_{t}$. A denoising U-Net is trained to recover the original structure of ${\mathbf z}_{t}$. At inference, ${\mathbf z}_{T}$ is initially sampled from a standard Gaussian distribution at timestep $\mathnormal T$ and is progressively denoised and restored to ${\mathbf z}_{0}$. A typical block in its U-Net includes three types of computations: 2D convolution, self-attention~\cite{attention_is_all_you_need}, and cross-attention. Cross-attention is conducted between condition $c$ and block features. 



\subsection{Learnable Gaussian Embedding}
\label{subsec:NGHP}
\subsubsection{Gaussian Representation}
We leverage 3D Gaussians~\cite{kerbl3Dgaussians} as geometric primitives to represent the head scene. These Gaussian primitives provide an adaptive and flexible representation of geometry, enabling more expressive control over head motion.
The Gaussian primitive is defined by a 3D covariance matrix $\mathbf{\Sigma}$ centered at a point $\mathbf{x_0}$:

\begin{equation} g(\mathbf{x}) = e^{-\frac{1}{2} (\mathbf{x}-\mathbf{x_0})^{T} \mathbf{\Sigma} ^{-1}(\mathbf{x}-\mathbf{x_0})}. \label{eq:gs} \end{equation}

The covariance matrix $\mathbf{\Sigma}$ can be decomposed into a rotation matrix $R$ and a scaling matrix $\varLambda$:

\begin{equation} \mathbf{\Sigma} = R\varLambda\varLambda^TR^T. \label{eq:cov} \end{equation}

We empirically found that using isotropic Gaussians as diffusion control signals yielded results similar to those obtained with anisotropic Gaussians. Therefore, we simplify the representation by setting $\varLambda = sI$. Since $RR^T = I$, the original quaternion $\mathbf{q}$ is no longer needed, leading to a simplified Gaussian formulation:

\begin{equation} g(\mathbf{x}) = e^{-\frac{1}{2s} \|\mathbf{x}-\mathbf{x_0}\|_{2}^{2}}. \label{eq:gs} \end{equation}

Each 3D Gaussian is additionally assigned two attributes: opacity $o$ and a learnable feature $\mathbf{f}$. The final simplified Gaussian can be written as $\{\mathbf{x_0}, s, o, \mathbf{f}\}$, while the entire Gaussian field is written as $(X_0, S, O, F)$.

\begin{figure}[tbp]
  \centering
  \includegraphics[width=\linewidth]{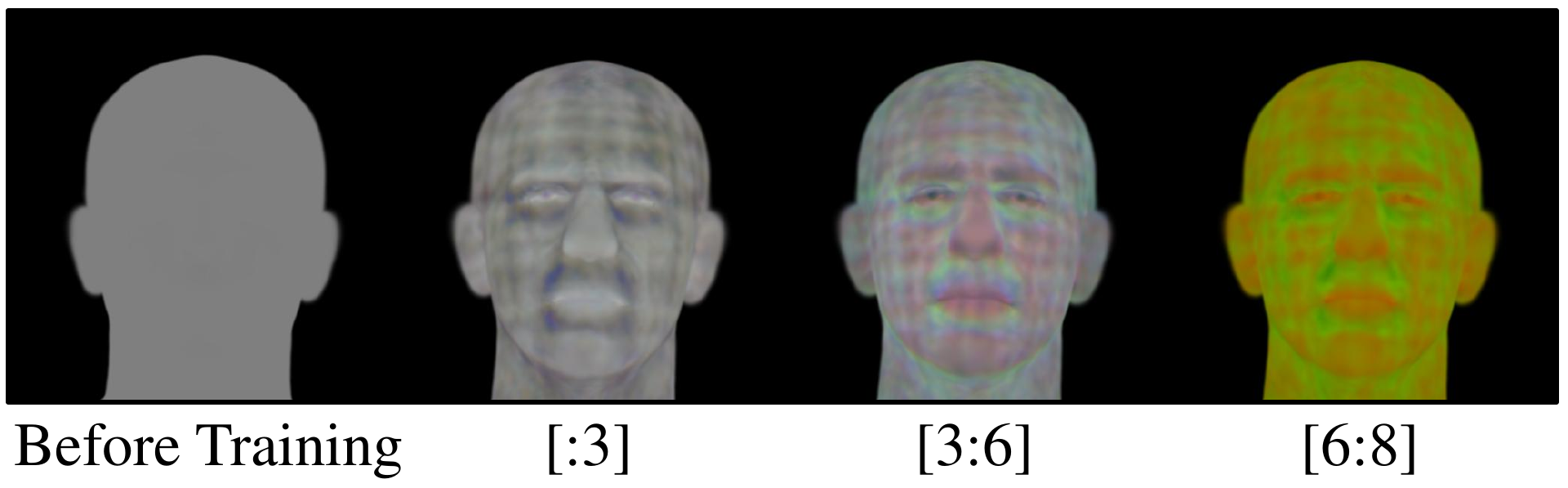}
  \caption{Visualization of the 8 channels of $\mathbf{F}$.  Before training, the feature maps exhibited no meaningful facial information. After training, they developed into highly expressive representations.}
  \label{fig:vis_fea}
  
\end{figure}

\subsubsection{Surface-embedded Gaussian Field}
We first track $\mathcal{I}_{tgt}$ to get parameters $ \mathcal{X}=\{(\beta,\theta_n,\psi_n,K_n,P_n)|n \in [1, N]\}$.
Here $\beta$, $\theta_n$, $\psi_n$ represent FLAME~\cite{FLAME:SiggraphAsia2017} shape, pose, and expression coefficients, while $K$,$P$ correspond to the camera's intrinsic and extrinsic parameters.

Similar to FlashAvatar~\cite{xiang2024flashavatar}, we maintain a 3D Gaussian field on the UV space of the FLAME model~\cite{FLAME:SiggraphAsia2017}, and further deform the Gaussians according to the deformation of underlying meshes. To be specific, we have a learnable Gaussian field $\phi$ in the FLAME UV domain where each pixel is characterized by three attributes: learnable feature $\mathbf{f}$, opacity $o$, scale $s$. Using UV mapping $\mathcal{G} \in \mathbb{R}^3 \rightarrow \mathbb{R}^2$, we transform the learnable Gaussians from UV space to 3D space.
This operation $\mathcal{H}$ could be written as:
\begin{equation}
    (X_0, S, O, F) = \mathcal{H}( M   \left(\beta,\theta,\psi\right) , \mathcal{G}, \phi),
\end{equation}
where $M   \left(\beta,\theta,\psi\right)$ is the FLAME surface.  Given $M   \left(\beta,\theta,\psi\right)$, $\mathcal{G}$ and $\phi$, we could get the 3D Gaussians $(X_0, S, O, F)$ corresponding to a certain frame. Then given camera's intrinsics $K$ and extrinsics $P$ ,  the splatted 2D feature map is computed by sorting and blending overlapped 3D Gaussians following ~\cite{kerbl3Dgaussians}:

\begin{equation}
  \mathbf{F} = \sum_{m} \mathbf{f}_m\alpha_m\prod_{j=1}^{m-1} (1-\alpha_j),
  \label{eq:blend}
\end{equation}

where $\alpha_m$ is obtained by multiplying the projected Gaussian with its opacity $o$. 

$\mathbf{F}$ is then used as control signal for the diffusion model in \Cref{subsec:diffusion}. Compared with previous works that utilize visual cues like landmarks, normal maps, or depth maps as head control signals, our proposed control signals are:
\begin{itemize}
\item \textbf{Dense} Unlike sparse landmark-based representations, the splatted feature map $\mathbf{F}$ provides a dense encoding of head shape and motion. This allows it to capture fine-grained deformations of the FLAME surface, such as subtle facial muscle movements.
\item \textbf{Adaptive} The joint optimization of learnable Gaussians and the denoising network enables the model to autonomously learn how to interpret local geometric features, significantly enhancing control accuracy.
\item \textbf{Expressive} Instead of storing colors or SH
 coefficients as Gaussian attributes, using a neural feature $\mathbf{f}$ make it possible to map the surface $M   \left(\beta,\theta,\psi\right) $ into a high-dimensional feature space, the overlapped Gaussians may exhibit more complex and expressive patterns.
\item \textbf{3D Consistent} It is worth noting that both normal maps and depth maps are not inherently 3D consistent, as their pixel values vary based on camera parameters. This further increases the learning burden of the denoising network and causes the inconsistency in final generated results. In contrast, our neural Gaussian head prior is a view-independent representation, which naturally encourage 3D consistency in the generated results.

\item \textbf{Easy for Cross-identity Reenactment} The FLAME model inherently possesses disentangled identity and expression spaces. For a cross-identity driving task, one only needs to replace the shape coefficients $\beta$ with the shape coefficients of reference image $\beta_{ref}$ to replace identity in control signals.

\end{itemize}

\subsection{Reference-Guided Denoising}
\label{subsec:diffusion}

With the learnable control signals proposed in \Cref{subsec:NGHP}, the parameter sequence $\mathcal{X}$ can be transformed into a feature map sequence $ \mathcal{C}=\{\mathbf{F}_n|n \in [1, N]\}$. In this subsection, we utilize a reference-guided denoising framework to learn the relationship between $\mathcal{C}$ and $\mathcal{I}_{tgt}$. Our framework contains a dual-branch paradigm including a reference branch and a denoising branch.

\subsubsection{Reference Branch}
Recent works have demonstrated that using a ReferenceNet~\cite{xu2023magicanimate,hu2024animate,ChangMagicPose} can significantly improve detail preservation, thanks to its multi-scale shared attention injection mechanism. In our implementation, ReferenceNet follows the same architecture as Stable Diffusion. Given the reference image $\mathbf{I}_{ref}$, we first encode the reference image with $\mathcal{E}$.
Then $\mathbf{z}_{ref}$ is passed through ReferenceNet. The feature maps in each transformer block of ReferenceNet are injected into the corresponding blocks of the denoising branch via shared attention.

\subsubsection{Denoising Branch}
Given the noised latent $\mathbf{z}_t$ and feature map sequence $\mathcal{C}$, the denoising network is trained to reconstruct the original structure of $\mathbf{z}_t$. We downsample $\mathcal{C}$ with the pose guidance network in~\cite{zhou2024realisdance} and add the feature to U-Net encoder. Every transformer block in denoising U-Net performs shared attention to integrate reference branch features into the denoising computation in a block-wise manner . Following previous video generation works~\cite{xu2023magicanimate,hu2024animate,ChangMagicPose}, temporal layers are applied to each block of the denoising U-Net to guarantee temporal consistency. 

\subsubsection{Loss function}
We use v-parameterization~\cite{salimans2022progressive} and the only loss function is:

\begin{equation}
  \mathcal{L}(\theta,\phi) = {\mathbb E}_{\mathbf{z}_0,t,\epsilon}(\|v_\theta(\mathbf{z}_t,\mathcal{C},\mathbf{z}_{ref})-v\|_2^2)
  \label{eq:loss}
\end{equation}

As illustrated in \Cref{fig:vis_fea}, there is no need to directly supervise $\mathbf{F}_n$. Instead, the Gaussian features $\mathbf{f}$  naturally develop into highly expressive representations under the guidance of the denoising loss function, highlighting the inherently adaptive nature of our control signal representation.

\subsection{Synthetic Head Dataset}
\label{subsec:synthetic}

Previous works primarily rely on talking head datasets for training, but most talking head videos contain limited pose or expression variations. As a result, models trained on these datasets lack robustness to large poses. Recent works have attempted to improve this by training the diffusion models using multi-view datasets or some internal large-pose datasets. However, the  number of subjects in these datasets is often limited (typically fewer than 1000), which is insufficient to train a model with strong identity generalization. In this work, we employ SphereHead~\cite{li2024spherehead} to generate a large-scale dataset with diverse poses and identities, significantly enhancing the pose robustness and identity generalization of reference-guided diffusion models. 

\begin{table}[tbp]
    \centering
    \begin{tabular}{lccc}
        \toprule
        & ID &  Views & In the wild \\
        \midrule
        D3DFACS~\cite{cosker2011facs} & 10  & 6 & NO \\
        NerSemble~\cite{kirschstein2023nersemble} & 268  & 16 & NO \\
        i3DMM~\cite{yenamandra2021i3dmm} & 64  & 137 & NO \\
        HUMBI~\cite{yu2020humbi} & 772  & 68 & NO \\
        Multiface~\cite{wuu2022multiface} & 13 & 160 & NO \\
        RenderMe~\cite{2023renderme360} & 500 & 60 & NO \\
        FaceScape~\cite{zhu2023facescape} & 847 & 68 & NO (with cap) \\
        MEAD~\cite{wang2020mead} & 48  & 7 & NO \\
        Our Synthetic Dataset & \textbf{10000} & \textbf{Dense} & \textbf{YES} \\
        \bottomrule
    \end{tabular}
    \caption{Comparison with existing multi-view head datasets}
    \label{table:dataset}
    \vspace{-5mm}
\end{table}

\begin{table*}[htbp]
    \begin{center}
        \resizebox{1.0\linewidth}{!}{
            \begin{tabular}{|c|ccccccc|ccc|}
                \hline
                \multicolumn{1}{|c|}{\multirow{2}{*}{Method}} &\multicolumn{7}{c|}{Self Reenactment} &\multicolumn{3}{c|}{Cross Reenactment}\\
                \cline{2-11}
                & PSNR $\uparrow$ & SSIM $\uparrow$ & LPIPS $\downarrow$ & L1 $\downarrow$ & CSIM $\uparrow$ & AED $\downarrow$ & APD $\downarrow$ & CSIM $\uparrow$ & AED $\downarrow$ & APD $\downarrow$ \\ 
                \cline{1-11}
                ROME~\cite{Khakhulin2022ROME} & 25.77 & 0.878 & 0.132 & 0.022 & 0.610 & 0.155 & \cellcolor{orange!30}0.016 & 0.426 & \cellcolor{orange!30}0.242 & \cellcolor{orange!30}0.025 \\ 
                Portrait4D-v2~\cite{deng2024portrait4dv2} & \cellcolor{orange!30}27.39 & \cellcolor{orange!30}0.887 & \cellcolor{yellow!30}0.104 & \cellcolor{orange!30}0.018 & 0.792 & \cellcolor{orange!30}0.135 & 0.022 & \cellcolor{yellow!30}0.584 & \cellcolor{yellow!30}0.247 & \cellcolor{yellow!30}0.028 \\ 
                VOODOO 3D~\cite{tran2024voodoo} & 22.10 & 0.856 & 0.152 & 0.034 & 0.499 & 0.171 & 0.045 & 0.357 & 0.250 & 0.054 \\ 
                GAGAvatar~\cite{chu2024gagavatar} & \cellcolor{yellow!30}26.58 & \cellcolor{yellow!30}0.886 & 0.117 & \cellcolor{yellow!30}0.019 & \cellcolor{yellow!30}0.810 & 0.143 & 0.030 & \cellcolor{orange!30}0.599 & \cellcolor{yellow!30}0.247 & 0.037 \\ 
                Follow-Your-Emoji~\cite{ma2024followyouremoji} & 26.39 & 0.878 & 0.111 & \cellcolor{yellow!30}0.019 & 0.792 & 0.176 & 0.023 & 0.432 & 0.388 & 0.118 \\ 
                X-Portrait~\cite{Xportrait} & 26.47 & 0.878 & \cellcolor{orange!30}0.103 & 0.020 & \cellcolor{orange!30}0.813 & \cellcolor{yellow!30}0.137 & \cellcolor{yellow!30}0.019 & 0.416 & 0.335 & 0.152 \\ 
                \cline{1-11}
                Ours & \cellcolor{red!30}27.84 & \cellcolor{red!30}0.890 & \cellcolor{red!30}0.099 & \cellcolor{red!30}0.016 & \cellcolor{red!30}0.831 & \cellcolor{red!30}0.123 & \cellcolor{red!30}0.014 & \cellcolor{red!30}0.648 & \cellcolor{red!30}0.232 & \cellcolor{red!30}0.023 \\ 
                \hline
            \end{tabular}}
        \caption{Quantitative results on the VFHQ~\cite{xie2022vfhq} testset.}
        \label{tab: quantitative_VFHQ}
    \end{center}
    \vspace{-3mm}
\end{table*}
\begin{table*}[htbp]
    \begin{center}
        \resizebox{1.0\linewidth}{!}{
            \begin{tabular}{|c|ccccccc|ccc|}
                \hline
                \multicolumn{1}{|c|}{\multirow{2}{*}{Method}} &\multicolumn{7}{c|}{Self Reenactment} &\multicolumn{3}{c|}{Cross Reenactment}\\
                \cline{2-11}
                & PSNR $\uparrow$ & SSIM $\uparrow$ & LPIPS $\downarrow$ & L1 $\downarrow$ & CSIM $\uparrow$ & AED $\downarrow$ & APD $\downarrow$ & CSIM $\uparrow$ & AED $\downarrow$ & APD $\downarrow$ \\ 
                \cline{1-11}
                ROME~\cite{Khakhulin2022ROME} & 24.06 & \cellcolor{yellow!30}0.892 & 0.155 & 0.030 & 0.331 & \cellcolor{orange!30}0.216 & \cellcolor{orange!30}0.025 & 0.424 & \cellcolor{orange!30}0.262 & \cellcolor{red!30}0.030 \\ 
                Portrait4D-v2~\cite{deng2024portrait4dv2} & \cellcolor{orange!30}25.56 & \cellcolor{yellow!30}0.892 & \cellcolor{orange!30}0.134 & \cellcolor{orange!30}0.024 & \cellcolor{orange!30}0.495 & \cellcolor{yellow!30}0.224 & \cellcolor{yellow!30}0.030 & \cellcolor{orange!30}0.589 & \cellcolor{yellow!30}0.271 & \cellcolor{orange!30}0.035 \\ 
                VOODOO 3D~\cite{tran2024voodoo} & 22.38 & 0.878 & 0.152 & 0.036 & 0.317 & 0.231 & 0.040 & 0.349 & 0.295 & 0.047 \\ 
                GAGAvatar~\cite{chu2024gagavatar} & \cellcolor{yellow!30}25.19 & \cellcolor{orange!30}0.902 & \cellcolor{yellow!30}0.144 & \cellcolor{yellow!30}0.025 & \cellcolor{yellow!30}0.450 & 0.231 & 0.041 & \cellcolor{yellow!30}0.575 & 0.286 & \cellcolor{yellow!30}0.046 \\ 
                Follow-Your-Emoji~\cite{ma2024followyouremoji} & 20.24 & 0.867 & 0.203 & 0.048 & 0.302 & 0.350 & 0.166 & 0.345 & 0.360 & 0.097 \\ 
                X-Portrait~\cite{Xportrait} & 19.75 & 0.845 & 0.200 & 0.051 & 0.229 & 0.351 & 0.274 & 0.341 & 0.376 & 0.191 \\ 
                \cline{1-11}
                Ours & \cellcolor{red!30}26.78 & \cellcolor{red!30}0.905 & \cellcolor{red!30}0.119 & \cellcolor{red!30}0.020 & \cellcolor{red!30}0.702 & \cellcolor{red!30}0.164 & \cellcolor{red!30}0.018 & \cellcolor{red!30}0.624 & \cellcolor{red!30}0.257 & \cellcolor{red!30}0.030 \\ 
                \hline
            \end{tabular}}
        \caption{Quantitative results on the NeRSemble~\cite{kirschstein2023nersemble} testset.}
        \label{tab: quantitative_NeRSemble}
    \end{center}
    \vspace{-7mm}
\end{table*}

\subsubsection{Dataset Generation}
Given a random noise input, SphereHead renders 3D-consistent head images of a certain subject. For each subject, we randomly sample with yaw angles in $\left[-60^{\circ}, 60^{\circ}\right]$ and pitch angles in $\left[{-30}^{\circ}, 45^{\circ}\right]
$, rendering approximately 60 images per subject. In total, we synthesize 10,000 unique subjects. We then fit FLAME models~\cite{FLAME:SiggraphAsia2017} to each subject’s multi-view images to estimate camera poses and FLAME coefficients.

\subsubsection{Mitigating Contamination}
Although synthetic images provide rich identity diversity, they often exhibit artifacts, particularly in regions such as ears, teeth, and eyeglasses. To prevent these artifacts from degrading the final generation quality, we introduce two learnable tokens to differentiate real and synthetic images. To be specific, during data loading, each image is assigned a label based on its origin: ``synthetic'' for synthetic images and ``real'' for images sourced from real videos. Each label corresponds to an optimizable vector, which is fed into the cross-attention module during the denoising process. Notably, only the ``real'' label is used during inference.
This strategy enables the model to effectively leverage the diversity of synthetic data while preserving the photorealism of real images. 

\subsubsection{Discussion}
As shown in \Cref{table:dataset}, our synthetic dataset contains a significantly larger number of subjects compared to existing multi-view datasets. Moreover, current datasets are captured under strictly controlled lighting conditions, making them unsuitable for training a generalizable head model. In contrast, SphereHead is trained on in-the-wild images, allowing it to reliably simulate diverse lighting conditions. 
Additionally, having full control over the viewpoints allows us to sample dense head poses, which is crucial for training data-hungry models like diffusion.

\section{Experiments}
\label{sec:experiments}

\begin{figure*}[ht]
  \centering
  \includegraphics[width=0.98\linewidth]{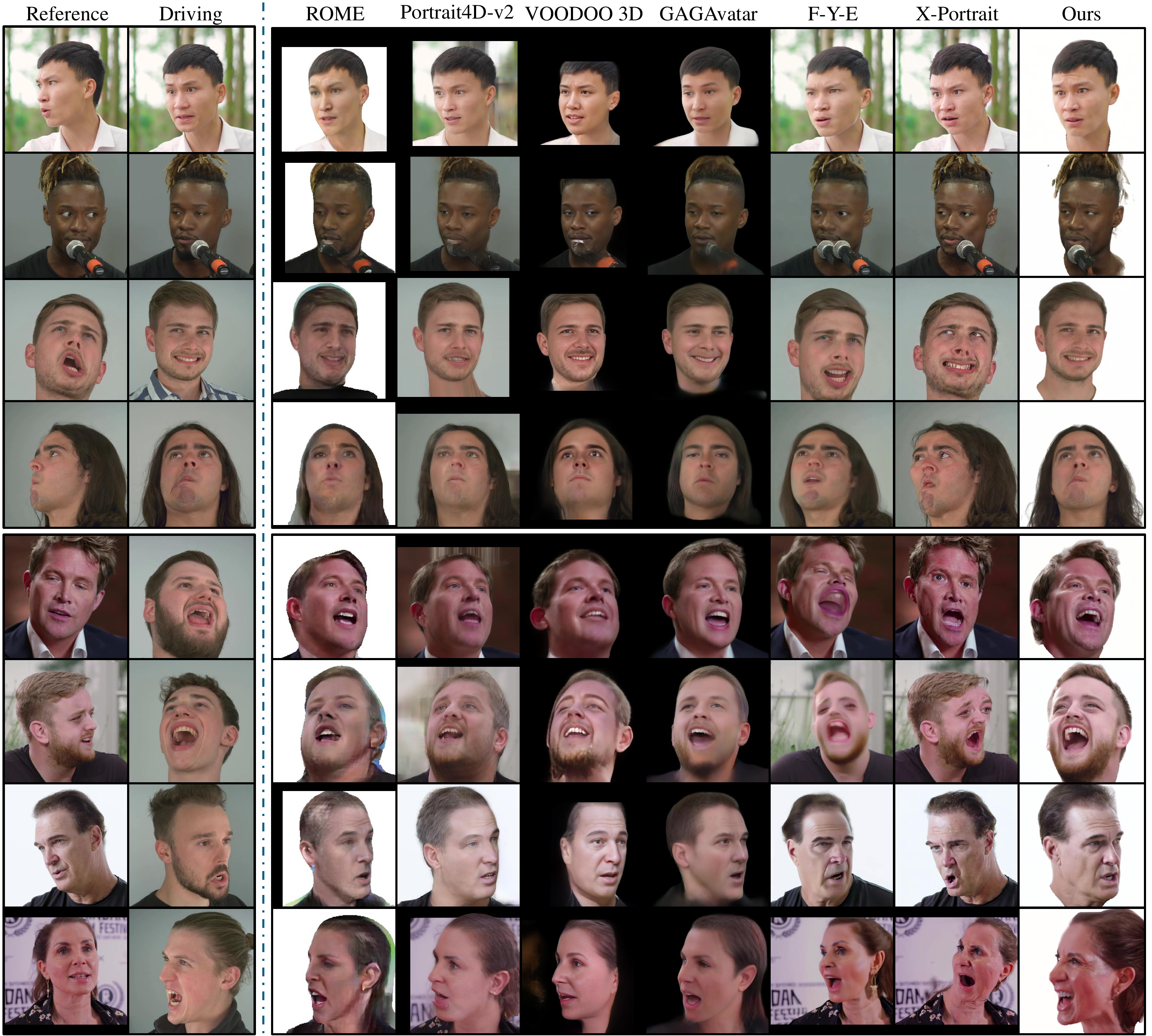}
  \caption{Comparison on the face reenactment task. We found that previous methods often struggled to preserve identity or expressions. In contrast, our approach not only faithfully reconstructs the identity from the reference but also maintains expression accuracy.}
  \label{fig:reenactment}
  \vspace{-3mm}

\end{figure*}



\begin{figure}[ht]
  \centering
  \includegraphics[width=\linewidth]{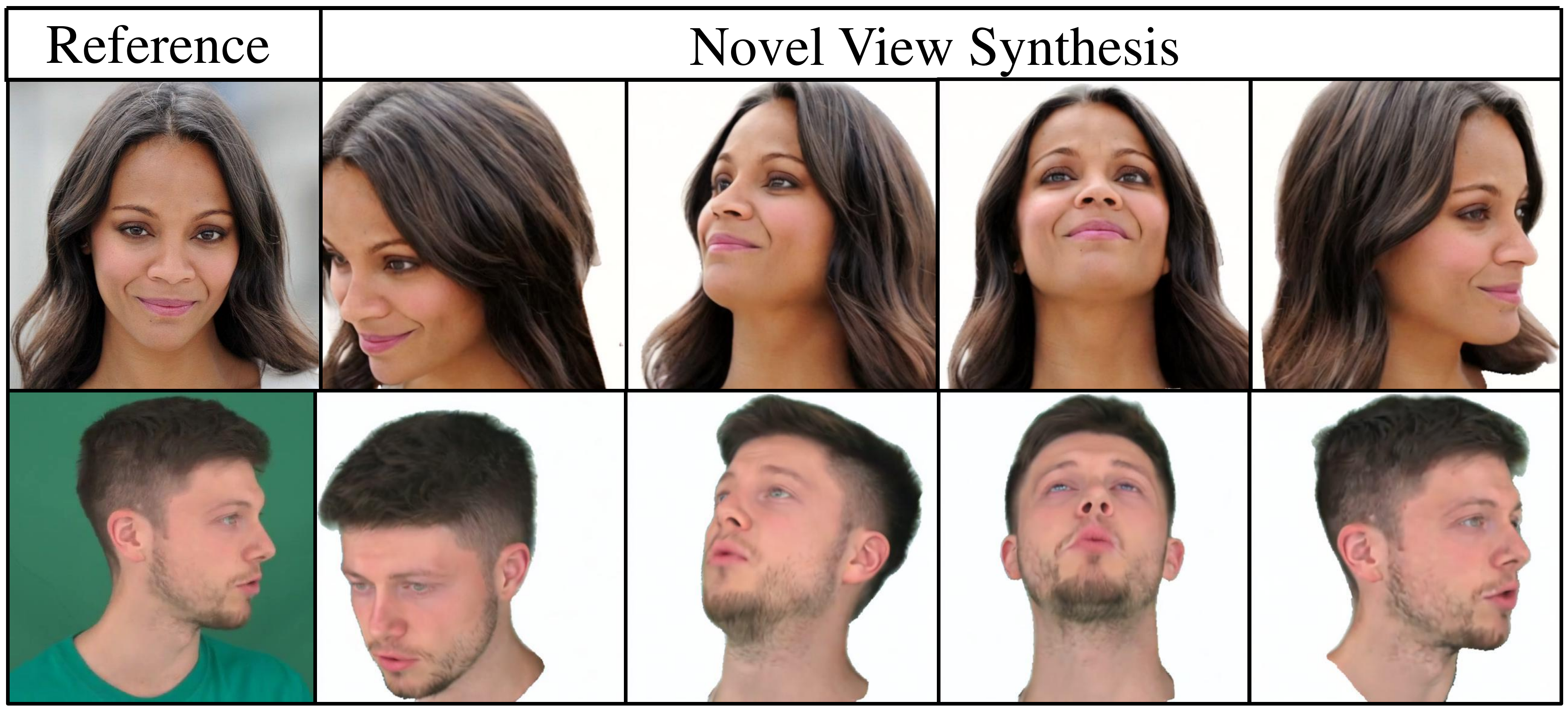}
  \caption{Novel view synthesis results. Given a single reference image, we can freely control the camera to synthesize view-consistent head images.}
  \label{fig:nvs}

\end{figure}
\subsection{Implementation Details}
We use VFHQ~\cite{xie2022vfhq}, NeRSemble~\cite{kirschstein2023nersemble}, MEAD~\cite{wang2020mead}, and our synthetic dataset to train our model. 
We adopt a two-stage training strategy to learn appearance modeling and temporal consistency separately. In the first stage, we set the sequence length to 1 and disable all temporal modules. We train the Gaussian attributes along with all remaining parameters in the dual branch. 
In the second stage, we increase the sequence length to 14. We train the temporal modules while keeping all other parameters fixed. The synthetic dataset is not used in this stage since 
multi-view frames are not continuous. All the frames are cropped and resized to $512\times512$ resolution. 

For the evaluation, we use two datasets to validate our methods, One is the VFHQ testset~\cite{xie2022vfhq}, which includes 50 talking videos. To further evaluate the motion accuracy in chanllenging cases, we use the last 20 subjects of NeRSemble~\cite{kirschstein2023nersemble} as another testset. For each subject we randomly select 2 sequences. Videos of these 20 subjects are excluded from the training dataset.



\subsection{Comparisons}
We compare our method with ROME~\cite{Khakhulin2022ROME}, Portrait4D-v2~\cite{deng2024portrait4dv2}, VOODOO 3D~\cite{tran2024voodoo}, GAGAvatar~\cite{chu2024gagavatar}, Follow-Your-Emoji (F-Y-E)~\cite{ma2024followyouremoji}, X-Portrait~\cite{Xportrait}.  ROME is a mesh based one-shot avatar. Portrait4D-v2 and VOODOO 3D are NeRF based head avatars. GAGAvatar is a 3D Gaussian based head model. Follow-Your-Emoji and  X-Portrait are based on video diffusion. We use the first frame as reference image for each sequence on VFHQ. For the NeRSemble testset, we randomly select one frame from the other video sequence of the same subject as the reference. Qualitative comparisons are shown in \Cref{fig:reenactment}. Our method could generate more expressive and consistent results compared with previous works. 
We found that when there is a significant difference in poses between the reference and the driving frame, most methods struggle to effectively preserve the identity information from the reference. ROME often failed to model hair region and facial details because of the limited representation ability of mesh model. We found that Portrait4D-v2, VOODOO 3D, and GAGAvatar often struggled to generate exaggerated expressions and maintain identity when dealing with significant pose deviations. This highlights the limited capacity of traditional one-to-one mapping head models. Follow-Your-Emoji and X-Portrait are trained with publicly available talking head datasets and several hundred internal video sequences. We observed that the results from both methods often exhibit severe facial distortions and inconsistencies. This further implies the data hungry nature of diffusion models and highlights the necessity of using our synthetic multi-view head dataset. 
Compared with these methods, our results exhibit much better consistency and motion accuracy, thanks to our learnable control signal representation and synthetic data utilization.

The quantitative results are shown in ~\Cref{tab: quantitative_VFHQ} and ~\Cref{tab: quantitative_NeRSemble}. For the self-reenactment task, we evaluate standard metrics such
as PSNR, SSIM, LPIPS~\cite{zhang2018perceptual}, and L1 distance. We measure identity preservation by comparing the cosine similarity between the embeddings of a
face recognition network~\cite{deng2019arcface} (CSIM). We compute Average Expression Distance (AED) and Average Pose Distance (APD) to evaluate the head motion accuracy. For fair comparison, we only compute these metrics in the head region. Our method outperforms previous works in all metrics,  which further verifies the effectiveness and superiority of our model.

\subsection{Novel View Synthesis}

To further evaluate the 3D consistency of our model, we conduct a novel view synthesis experiment. Given a single reference image, we manipulate the poses of the generated head images by adjusting the pose parameters of the FLAME head model. As illustrated in \Cref{fig:nvs}, our method produces reasonable and consistent results even for large pose variations. This demonstrates that our learnable Gaussian embedding, combined with training on a synthetic dataset, effectively enhances the 3D consistency of diffusion models.

\section{Ablation Study}
\label{sec:ablation}
\begin{figure}[tbp]
  \centering
  \includegraphics[width=\linewidth]{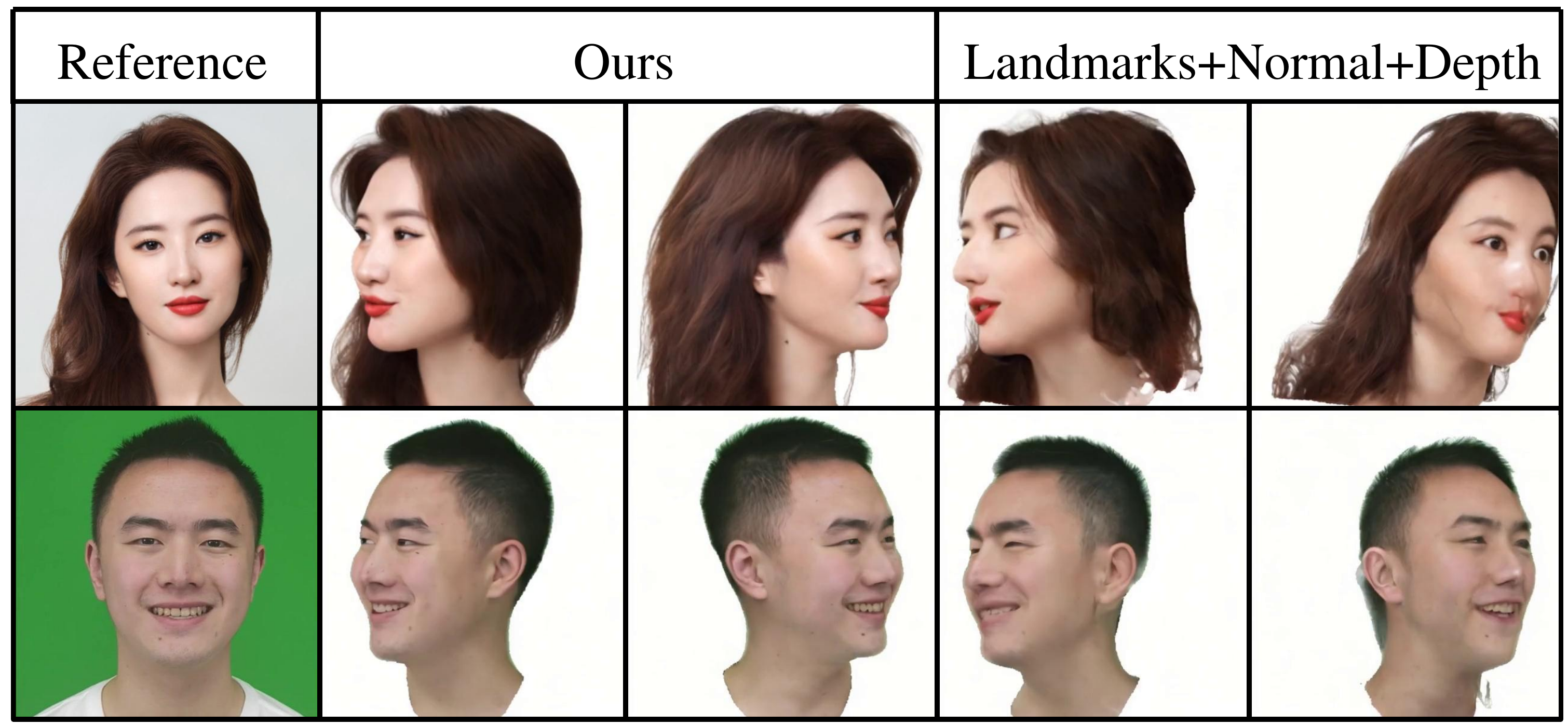}
  \caption{The figure illustrates the head of the reference image at a yaw angle of $\pm
50^{\circ}$ generated by our model and the baseline diffusion model (controlled by traditional visual cues).}
  \label{fig:abl_control}
\end{figure}

\begin{figure}[ht]
  \centering
  \includegraphics[width=\linewidth]{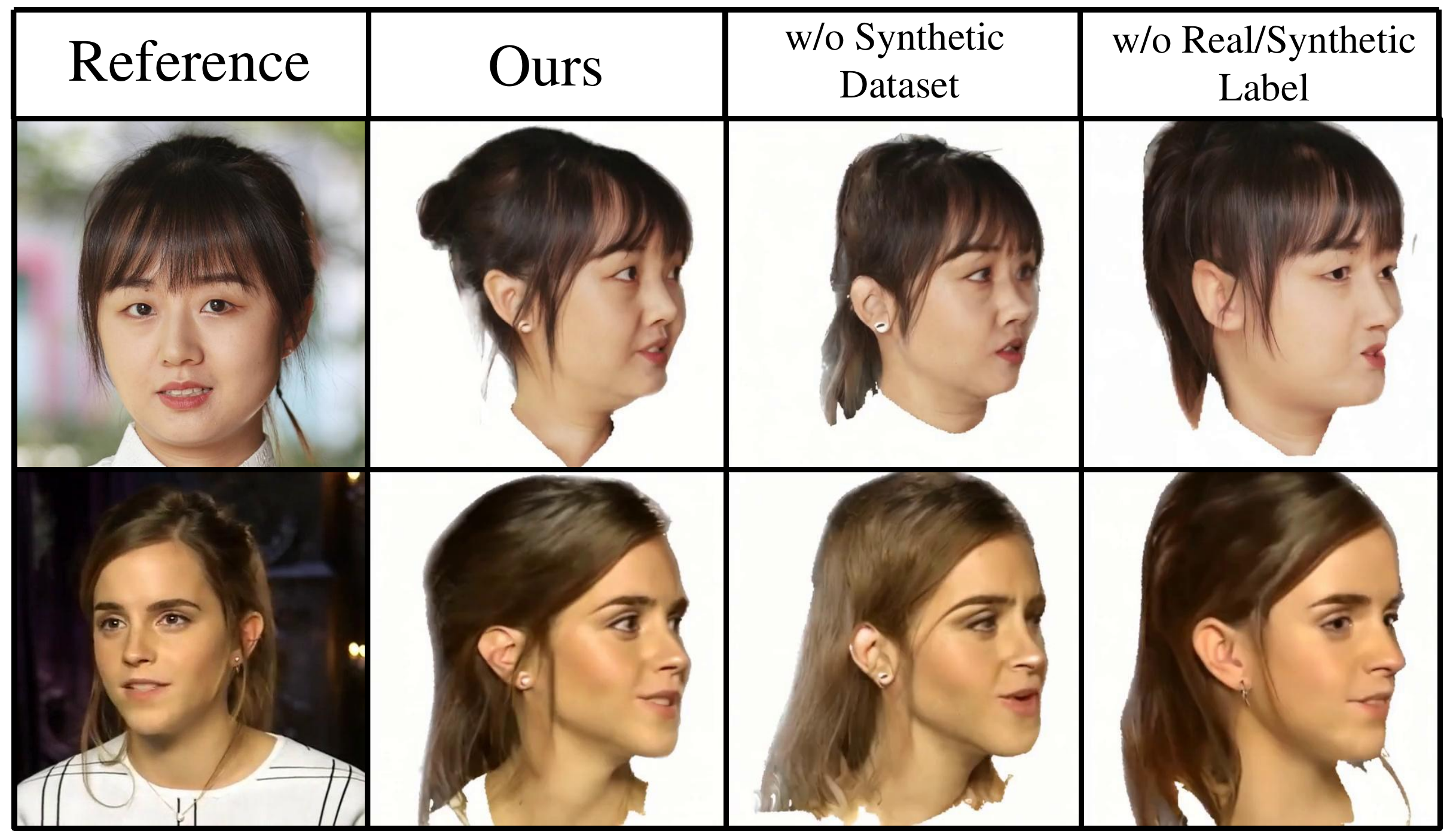}
  \caption{Without training on our synthetic dataset, the model may fail to synthesize head images in large poses . And if the model is trained without Real/Synthetic labels, the artifacts of the synthesized samples may influence the generated results.}
  \label{fig:abl_data}
  \vspace{-3mm}
\end{figure}
\subsection{Learnable Gaussian Embedding}
Some previous diffusion based portrait generation models employ landmarks, normal maps, or depth maps as signals to control diffusion generation. To demonstrate the deficiencies of these visual cues, we design a baseline model that concatenate FLAME's landmarks, normal maps, depth maps as control signals. As shown in ~\Cref{fig:abl_control}, such a model fails to accurately generate images that meet the required expressions and poses.
This primarily stems from the sparsity of landmarks, as well as the low-frequency nature and 3D inconsistency of normal and depth information. Our learnable Gaussian feature map $\mathbf{F}$ is a dense, adaptive, expressive, and 3D consistent control signal representation. It showcased better quality in controlling head motion generation.

\subsection{Synthetic Dataset}

We utilize a synthetic dataset to enhance the model's identity generalization and view robustness. Additionally, we incorporate real/synthetic labels to mitigate the impact of artifacts in synthetic data on the generated results. As illustrated in ~\Cref{fig:abl_data}, models trained solely on public real datasets struggle to synthesize head images with large poses. Moreover, without our real/synthetic label as input, noticeable artifacts may appear in the generated results.
\section{Conclusion and Discussion}
\label{sec:conclusion}

In this paper, we enhanced the generalization and 3D consistency of head diffusion through two key innovations: control signal representation and dataset design. First, we introduced splatting-based feature maps derived from learnable Gaussian embeddings, which serve as dense, adaptive, and 3D-consistent control signals. Second, we incorporated synthetic data to improve consistency and generalization, paired with a real/synthetic labeling strategy to mitigate artifacts. Experiments demonstrated that these contributions collectively advance the fidelity and robustness of head generation.

Notably, the framework of augmenting diffusion models with learnable Gaussian embeddings is not restricted to head avatars. We envision extensions to human body animation or generalizable video generation, where 3D-aware control signals could similarly address consistency challenges. We leave these promising directions for future work.

\noindent\textbf{Acknowledgements.} This research was supported by the National Natural Science Foundation of China (No.62441224, No.62272433). The numerical calculations in this paper have been done on the supercomputing system in the Supercomputing Center of University of Science and Technology of China.
{
    \small
    \bibliographystyle{ieeenat_fullname}
    \bibliography{main}
}

\clearpage
\setcounter{page}{1}
\maketitlesupplementary
\appendix

\section{Dataset Details}
We use VFHQ~\cite{xie2022vfhq}, NeRSemble~\cite{kirschstein2023nersemble}, MEAD~\cite{wang2020mead}, and our synthetic dataset to train our model. 

\noindent
\textbf{VFHQ} VFHQ is a talking video dataset containing approximately 16,000 subjects. While most videos in this dataset are of high quality, the head movement within each sequence is relatively limited.

\noindent
\textbf{NeRSemble} NeRSemble consists of 16 viewpoints and 268 subjects. It recorded many exaggerated expressions. We randomly sampled about 4000 sequences from this dataset for training. The last 20 subjects are excluded from training dataset and left for evaluation.

\noindent
\textbf{MEAD} MEAD consists of 7 viewpoints and 48 subjects. We randomly sampled about 2000 sequences from this dataset for training.


\section{Implementation Details}
We use a model similar to~\cite{retsinas20243d} to track the video datasets.  For the synthetic multi-view dataset, we get the FLAME parameters and camera parameters through multi-view landmarks fitting. The dataset sampling probabilities (VFHQ, NeRSemble, MEAD, synthetic) are set to 1:0.8:0.2:1 for the first stage and 1:0.8:0.2:0 for the second stage. The UV resolution of the Gaussians is $256\times256$. The latent channels of $\mathbf{F}$ is 8. We use the pose guidance network in~\cite{zhou2024realisdance} to encode our learnable control signals $\mathbf{F}$ into U-Net. The pose guidance network accepts the signals as single modal with channels 8. $tanh$ is used to make the features in $[-1,1]$.

\begin{figure*}[ht]
  \centering
  \includegraphics[width=\linewidth]{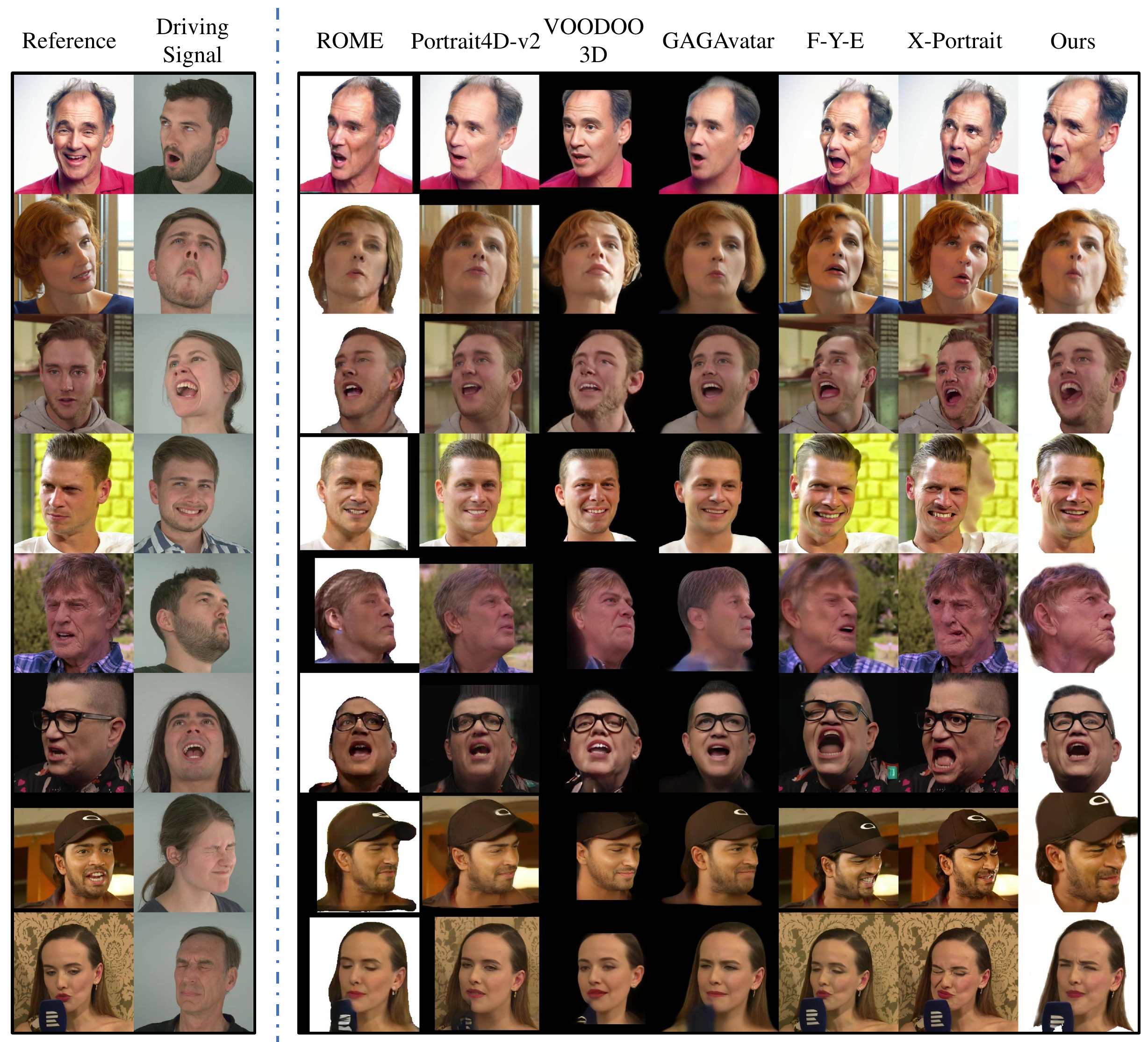}
  \caption{More results on cross-identity reenactment}
  \label{fig:cross_large}

\end{figure*}



\section{Evaluation Details}
\subsection{Metrics}
\paragraph{LPIPS.} We use the VGG based perceptual similarity metric LPIPS~\cite{zhang2018perceptual} to measure the perceptual similarity between the animated and the driving images.

\paragraph{AED.} AED is the mean $\mathcal{L}_1$ distance of the expression parameters between the animated and the driving images. These parameters, which include facial movement, eyelid, and jaw pose parameters, are extracted by the state-of-the-art 3D face reconstruction method SMIRK~\cite{retsinas20243d}.


\paragraph{APD.} APD is the mean $\mathcal{L}_1$ distance of the pose parameters between the animated and the driving images. The pose parameters are extracted by SMIRK~\cite{retsinas20243d}.

\paragraph{CSIM.} CSIM measures the identity preservation between two images, through the cosine similarity of two embeddings from a pretrained face recognition network~\cite{deng2019arcface}. For self-reenactment, the CSIM is calculated between the animated and the driving images. For cross-reenactment, the CSIM is calculated between the animated and the source portraits.





\section{More Results on Cross-identity Reenactment}
We show more cross-identity reenactment results in \Cref{fig:cross_large}

\section{Limitations}
Although our works have showcased remarkable improvements compared with previous works, it still have some limitations. Our method still relies on a head tracking algorithm to get the control signals. Therefore, large errors in tracking, especially global pose errors, may cause misalignment. The bias of the datasets may also influence the generated results.

\section{Broader Impact}

Given a single reference image, our model generate consistent and expressive head animation in arbitrary expressions and poses. 
However, there is a risk of misuse, \eg the so-called DeepFakes. We strongly discourage using our work to generate fake images or videos of individuals with the intent of spreading false information or damaging their reputations. Unfortunately, we may be unable to prevent the nefarious use of our technology. Nevertheless, we believe that performing research in an open and transparent way could raise the public's awareness of nefarious uses, and our work could further enhance forgery detection capabilities.

\end{document}